# 350-GHz-Band 4 × 4 RTD Monostatic Radar Array for Sequential Multidirectional Ranging

Li Yi[1,4] Member, IEEE, Ryoma Nakamura[1], Shota Ito[1], Yousuke Nishida[2], Koji Terumoto[2], Toshihisa Maeda[2], Bryce Chung[3], Student Member, IEEE, Yunbin Jiang[3], and Daniel Headland[3], Member, IEEE
[1] Graduate School of Science and Engineering, Ibaraki University, Hitachi, Japan
[2] ROHM Research & Development Center, ROHM Co., Ltd., Kyoto, Japan
[3] Terahertz Engineering Laboratory, Adelaide University, Adelaide, Australia
[4] Department of Electrical Engineering and Information Systems, The University of Tokyo, Tokyo, Japan

*Abstract*—Terahertz (THz) sensing offers significant potential for nondestructive evaluation, imaging, and metrology, but the cost and complexity of existing systems remain barriers to practical deployment. This letter presents a compact 4 × 4 monostatic radar array based on a single-RTD-per-pixel architecture, in which each resonant tunneling diode (RTD) functions as both a bias-tunable oscillator and a self-mixing detector. The RTD elements are sequentially addressed through a low-frequency switching network and share the same baseband control and readout electronics. A shared 3D-printed dielectric lens maps the RTD elements to spatially separated sensing directions. We experimentally verify the operation of all array elements and demonstrate sequential multidirectional ranging using four selected elements, followed by preliminary 16-pixel THz imaging. By combining self-mixing at each pixel with low-frequency selection and shared optics, the array avoids a separate THz receiver chain for each pixel and provides a compact architecture for electronically addressable THz sensing and imaging.

***Index Terms*—Resonant tunneling diode (RTD), terahertz sensing, monostatic radar array, self-mixing radar, terahertz imaging.**

## I. Introduction

Terahertz (THz) waves (~0.1–10 THz) combine submillimeter wavelengths, material-dependent dielectric responses, and the ability to penetrate many common nonconducting materials. These characteristics enable applications in nondestructive evaluation, imaging, spectroscopy, and metrology [1]–[4]. In addition, the short wavelengths support highly integrated sensor architectures that can be combined with compact optical or electronic imaging systems, creating opportunities for multimodal sensing in robotic perception and autonomous systems.

Among established THz techniques, terahertz time-domain spectroscopy (THz-TDS) enables broadband permittivity extraction and time-of-flight-based thickness or distance estimation [3], [4]. However, its reliance on femtosecond lasers and complex optoelectronic components often results in bulky, costly, and power-intensive systems, which conflict with low size, weight, and power (SWaP) requirements. THz radar provides a more compact alternative when low power consumption and simplified electronics are priorities [2] – [4]. Recent CMOS-based submillimeter-wave radar and imaging systems have advanced integrated active sensing [5] – [7]; however, operation near and above 300 GHz remains constrained by limited output power, front-end losses, and the complexity of implementing transmitter and receiver functions at each pixel.

Resonant tunneling diodes (RTDs) are attractive for compact THz sensing because they support room-temperature THz oscillation, voltage-controlled frequency tuning, and strong intrinsic nonlinearity [8]–[11]. RTD-based radar and self-mixing sensors have demonstrated coherent reception, frequency-modulated ranging, displacement sensing, and thickness measurements [11]–[13]. A single RTD can therefore function as both a tunable source and a self-mixing detector, providing a compact monostatic sensing element.

In this work, we extend the single-RTD monostatic concept to a 4 × 4, 350-GHz-band radar array. Because the closely spaced on-chip dipole antennas have limited directivity and effective aperture, a shared 3D-printed dielectric lens [14] is used to enlarge the effective radiation aperture and map each array element to a spatially separated sensing direction. The architecture combines three features: each pixel uses one RTD as both a transmitter and a self-mixing detector; the pixels are sequentially addressed through a low-frequency switching network instead of duplicated THz receiver chains; and the lens converts the compact array into an electronically selectable multidirectional sensor. This differs from CMOS concurrent-transceiver arrays that integrate transmitter, receiver, and local-oscillator circuitry within each pixel [7], RTD arrays developed for source power combining [15], and lens-fed THz switches used for beam selection [16]. The experiments focus on array-level functionality: all 16 RTDs are individually addressed, four selected elements are used for multidirectional ranging, and a second-lens configuration provides preliminary 16-pixel imaging.

## II. THz Monostatic Array Using RTDs

### *A.* THz Radar Enabled by a Single RTD

The chip used in this work is illustrated in Fig. 1(a). It consists of a single RTD coupled directly to a planar on-chip dipole antenna. The bias circuit is connected to the diode through an RF-blocking structure comprising two metal layers separated by a thin insulator. Between the RF block and the RTD, a short slot line section forms a cavity resonator that serves as the tuning circuit for the THz integrated circuit [9].

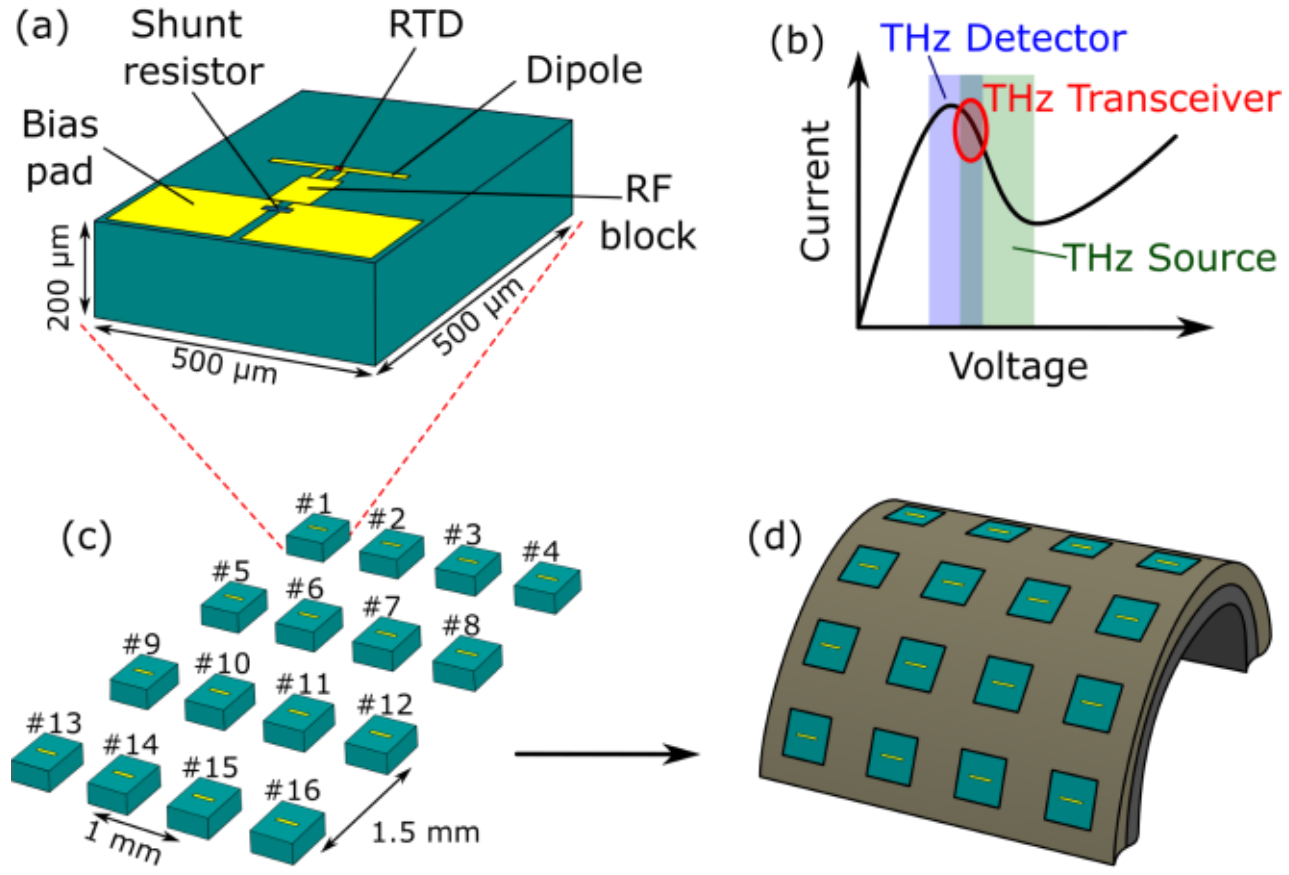

Fig. 1. (a) Structure of the RTD chip, adapted from [9]. (b) RTD operating modes at different bias voltages, including the transition region in which simultaneous oscillation and mixing can occur. (c) Array of discrete single-diode chips investigated in this work. (d) Conceptual conformal arrangement, not experimentally evaluated in this work.

The sole active component in the chip shown in Fig. 1(a) is an RTD, a quantum heterostructure device that exhibits negative differential conductance (NDC) in its current–voltage characteristic [8], [9], as shown in Fig. 1(b). At a suitable bias point, the NDC and resonator sustain oscillation in the 350-GHz band. The diode nonlinearity also supports coherent mixing [11]. In self-mixing operation, the returned THz reflections interacts with the existing oscillation and changes the terminal response [10], [13], allowing the same RTD to act as both a source and a detector.

When biased in the intermediate region indicated in Fig. 1(b), the RTD simultaneously operates as a voltage-tunable oscillator and a nonlinear mixer [13]. For an ideal linear frequency sweep, the transmitted frequency is expressed as

$$f_{\mathrm{TX}}(t) = f_0 + Kt, \qquad (1)$$

where $f_0$ is the initial oscillation frequency and $K$ is the chirp rate. The signal reflected from a stationary target at range R returns to the RTD after a round-trip delay $\tau$. Owing to the nonlinear current-voltage characteristic of the RTD, the delayed echo is coherently mixed with the instantaneous oscillation in the same device, producing a baseband self-mixing signal

$$v_{\mathrm{BB}}(t) = A\cos\ (2\pi f_{\mathrm{b}} t + \phi_{\mathrm{b}}), \qquad (2)$$

where $A$ and $\phi_{\mathrm{b}}$ denote the amplitude and phase of the baseband response, respectively. For an ideal linear chirp and a stationary target, the beat frequency is

$$f_{\mathrm{b}} = K\tau = \frac{2KR}{c}, \qquad (3)$$

allowing the target range to be obtained from the measured beat frequency. Thus, the same RTD radiates the frequency-swept THz signal, coherently mixes the returned echo, and generates a directly measurable baseband beat signal on its bias line, thereby realizing a compact single-diode monostatic frequency-modulated continuous-wave (FMCW) radar, as shown in Fig. 2(b).

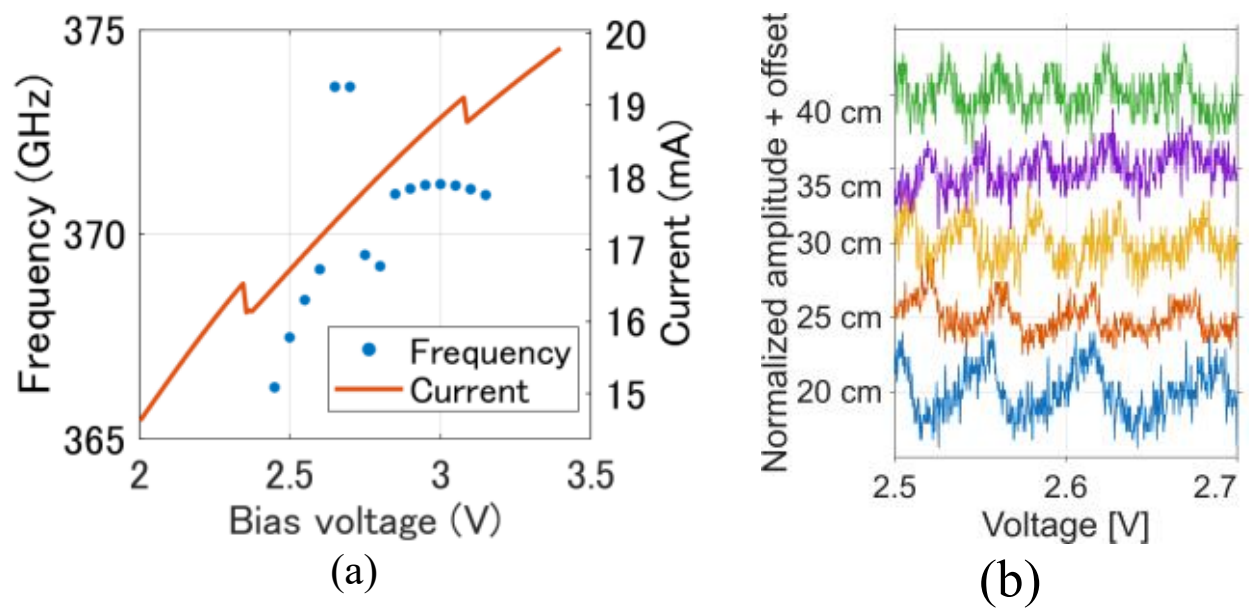

Fig. 2. (a) Bias-dependent emission frequency and current of RTD #15. When the bias voltage was chirped from 2.5 to 2.7 V, the emission frequency was swept over a bandwidth of approximately 5 GHz. (b) Corresponding self-mixing signals for metallic plates at distances of 20–40 cm; the beat frequency increases with target distance.

The RTD self-mixing signal approximates the radar beat signal, as discussed in [13]. Feedback-induced frequency pulling and nonlinear voltage-to-frequency tuning can distort the waveform [10], so element-specific calibration is used in Section III.

High-frequency mixing occurs intrinsically within the RTD chip, whereas frequency tuning by bias modulation and self-mixing readout are implemented at baseband [13]. In our implementation, a field-programmable gate array (FPGA) supplies the bias waveform through a digital-to-analog converter (DAC), and an analog-to-digital converter (ADC) records the output of a differential amplifier connected to the bias line. This common baseband interface allows the selected pixel to be connected through low-frequency switches rather than providing a separate THz receiver chain for each array element, as illustrated in Fig. 3.

In this work, we extend this monostatic architecture to a 4 × 4 array, as illustrated in Fig. 1(c), to demonstrate its scalability. Because the chips are physically discrete, the architecture may also support irregular or conformal geometries, as illustrated conceptually in Fig. 1(d); such configurations are not evaluated in the present experiment.

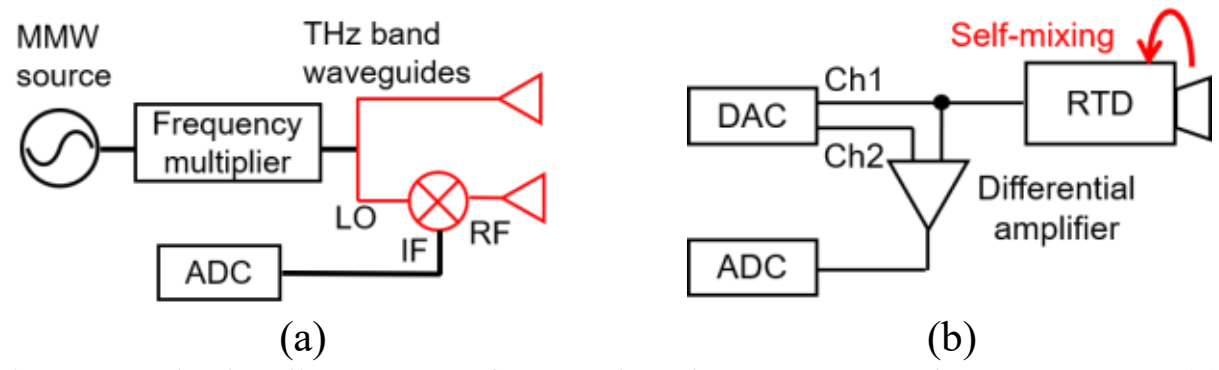

Fig. 3. Block diagrams of THz-band FMCW radar systems. (a) Conventional system using frequency multiplier and a mixer. (b) Proposed control circuit in which a single RTD operates as both a self-oscillating mixer and a tunable oscillator.

### *B.* 350-GHz-Band Monostatic Array

In this work, 16 RTD chips were assembled into a 4 × 4 array. Each RTD chip occupies an area of approximately 500 μm × 500 μm. The edge-to-edge gaps are 1.0 and 0.5 mm along the two array directions, corresponding to center-to-center pitches of 1.5 and 1.0 mm. As shown in Fig. 4, low-frequency switches sequentially connect the selected RTD pixel to the common bias and readout circuit.

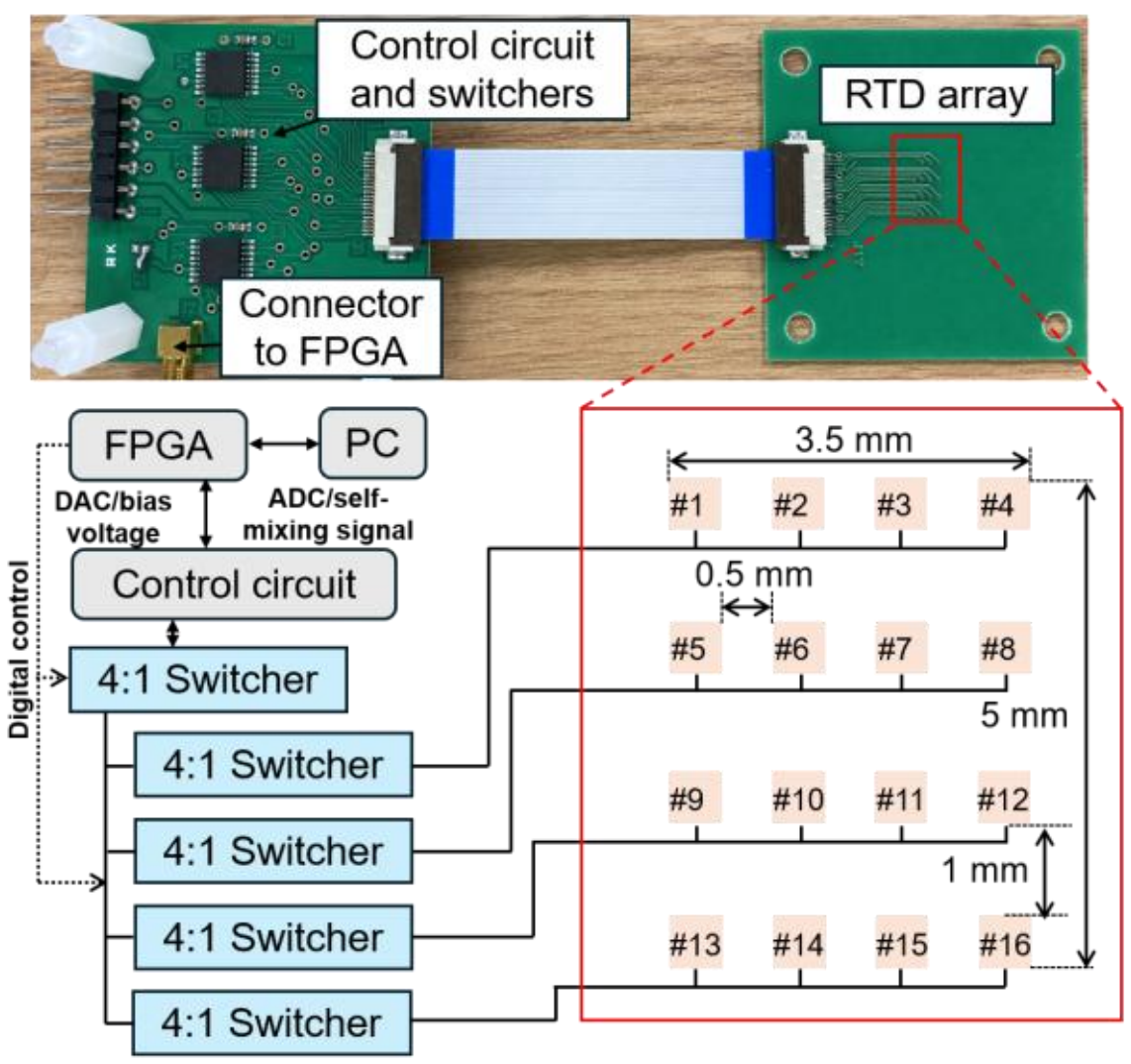

Fig. 4. Architecture of the 4 × 4 RTD monostatic array. The FPGA generates the bias waveform through a DAC and sequentially addresses the 16 RTD pixels through a low-frequency switching network. The self-mixing signal from the selected pixel is routed to the ADC for digitization.

The 16 RTDs are controlled by an FPGA board and operated sequentially on a pixel-by-pixel basis. A 1-kHz bias chirp is applied to the selected RTD to enable the FMCW operation described in Section II-A. Ideally, one scan of all 16 pixels requires 16 ms. In practice, additional time is required for switching, low-pass filtering, Fourier transform, and subsequent radar signal processing. Under the current acquisition settings, the measurement time is approximately 0.1 s per pixel.

Because the RTD chips were mounted manually, small placement deviations were unavoidable. Device-to-device variations also resulted in different operating-frequency bands and required different bias voltages. In the present array, 9 RTDs operate in the 370-GHz band, whereas the remaining 7 RTDs operate in the 330-GHz band. The useful self-mixing frequency excursion depends on the selected device and bias interval; the values for the four ranging pixels are given in Section III-A. Each pixel is therefore operated at a suitable bias and calibrated individually. Sequential operation does not require mutual phase locking between pixels.

To evaluate the radiation characteristics of the array, the near-field distributions of all 16 RTD elements were measured using a spectrum analyzer equipped with a frequency extender. For each pixel, the radiated signal was sampled over a 1-cm-wide region at a spatial interval of 1 mm, as illustrated in Fig. 5. The results confirm that all 16 RTDs operate and that each pixel can be addressed individually. The measured power was below 1 μW at a probe distance of approximately 1 cm, highlighting the need for quasi-optical aperture enhancement to improve the signal-to-noise ratio (SNR).

### *C.* Customized Telecentric Lens for the RTD Array

The compact integrated dipoles produce broad radiation patterns. To improve the SNR, a shared 3D-printed dielectric lens is introduced to enlarge the effective aperture and map different RTD positions to spatially separated sensing directions. A low-loss dielectric material is required to limit absorption. Cyclic olefin copolymer (COC) exhibits low loss in the THz range and can be processed using consumer-grade 3D printers, making it suitable for customized THz quasi-optics [14], [17].

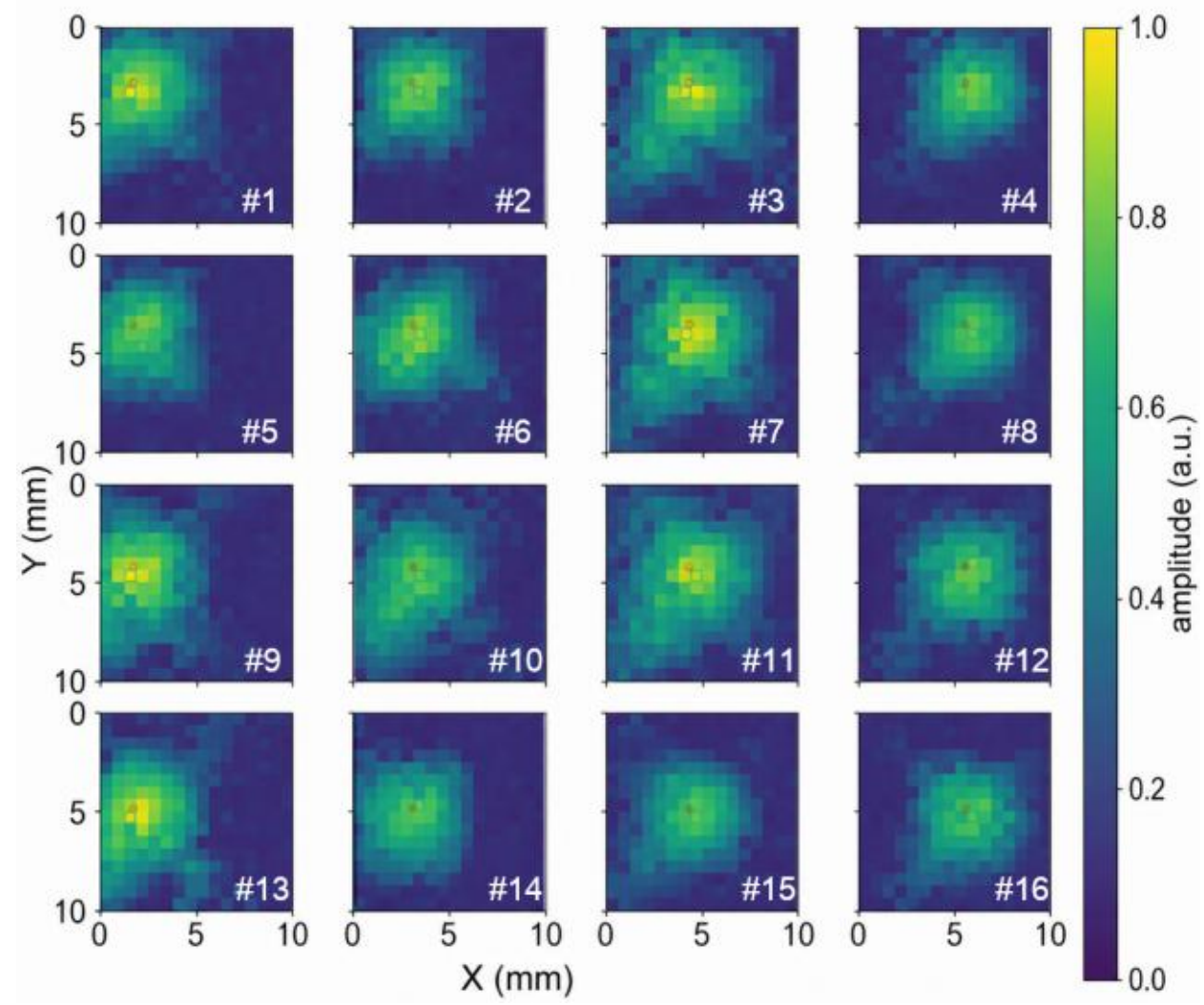

Fig. 5. Measured near-field radiation distributions of the 16 individually addressed RTD elements. The measurement plane was located approximately 1 cm from the array. Each map is normalized separately and therefore does not provide an absolute-power comparison between pixels.

To accommodate the broad radiation pattern of the on-chip dipole feed, a focal ratio of $F/D = 1$ is selected. Accordingly, a lens diameter of 30 mm and a focal length of 30 mm are adopted to maintain a compact system footprint. Ansys Zemax OpticStudio is used to optimize the lens curvature for the array geometry following the telecentric optical approach in [14]. The single-lens configuration is used for multidirectional sensing, as shown in Fig. 6(a), whereas a second lens projects the array onto a target plane for 2D imaging, as shown in Fig. 6(b).

Because all RTD elements share the same customized lens, separate lens-based collimation and alignment for individual pixels are avoided. The 3D-printed aspheric lens can also be adapted to different RTD array configurations.

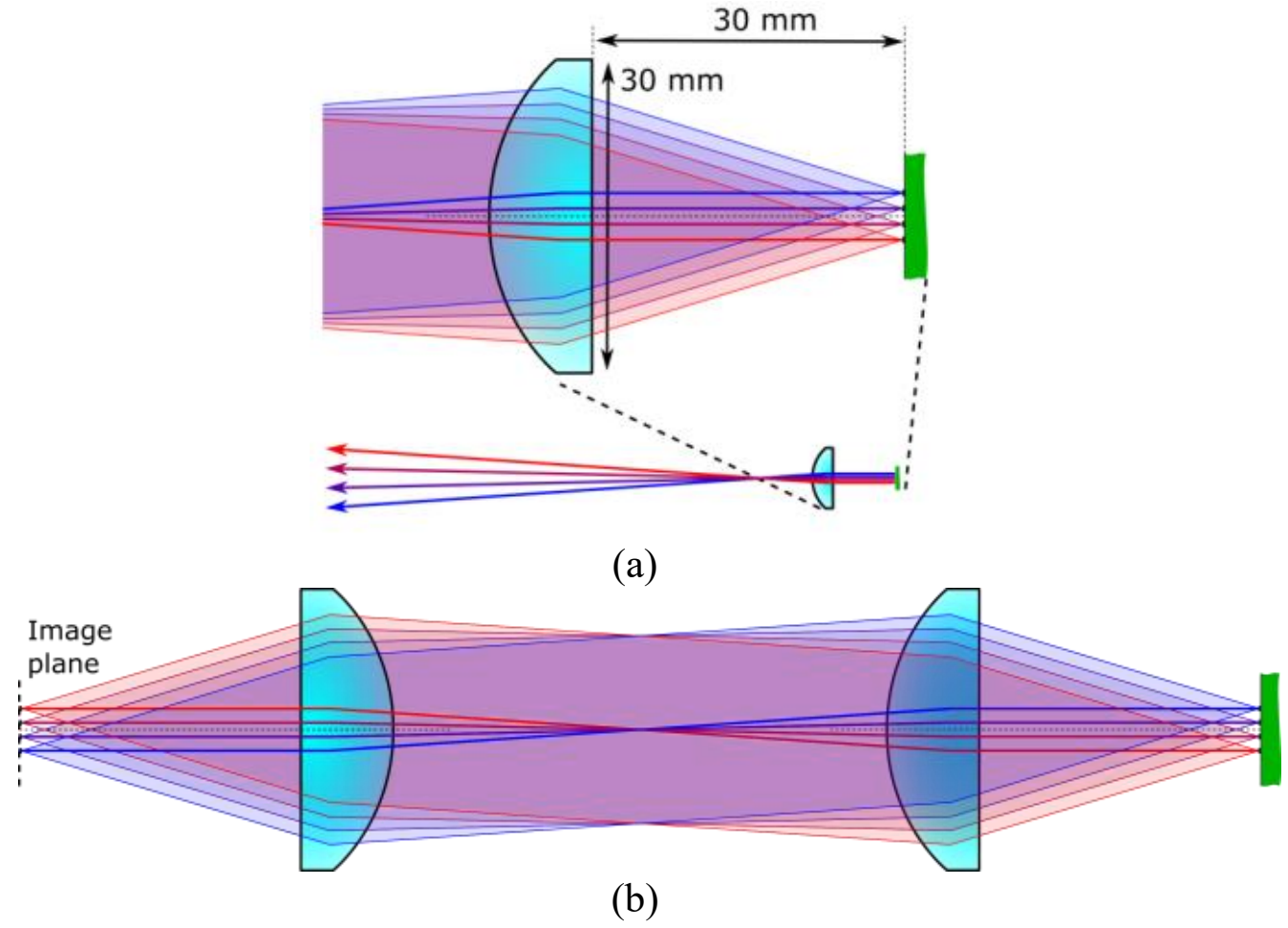

Fig. 6. Configurations of the 3D-printed telecentric lens. (a) Single-lens configuration for multidirectional sensing. (b) Two-lens configuration for projecting the array onto the imaging plane.

## III. Experimental Results

### A. Sequential Multidirectional Terahertz Ranging Using the RTD Array

As a proof-of-concept demonstration of multidirectional THz radar operation, RTDs #3, #7, #11, and #15 were sequentially addressed to probe four spatially separated directions. Their bias voltages were adjusted within 2.5–2.7 V to achieve stable oscillation, giving effective frequency excursions of approximately 10, 9, 8, and 5 GHz, respectively. The shared lens maps the positions of the selected RTDs to different target regions. With a focal length of 30 mm, the angular separation between adjacent directions was approximately 3°, as shown in Fig. 7. Three handmade corner reflectors were placed at reference ranges of 40, 50, and 45 cm for RTDs #3, #7, and #15, respectively; no reflector was assigned to RTD #11.

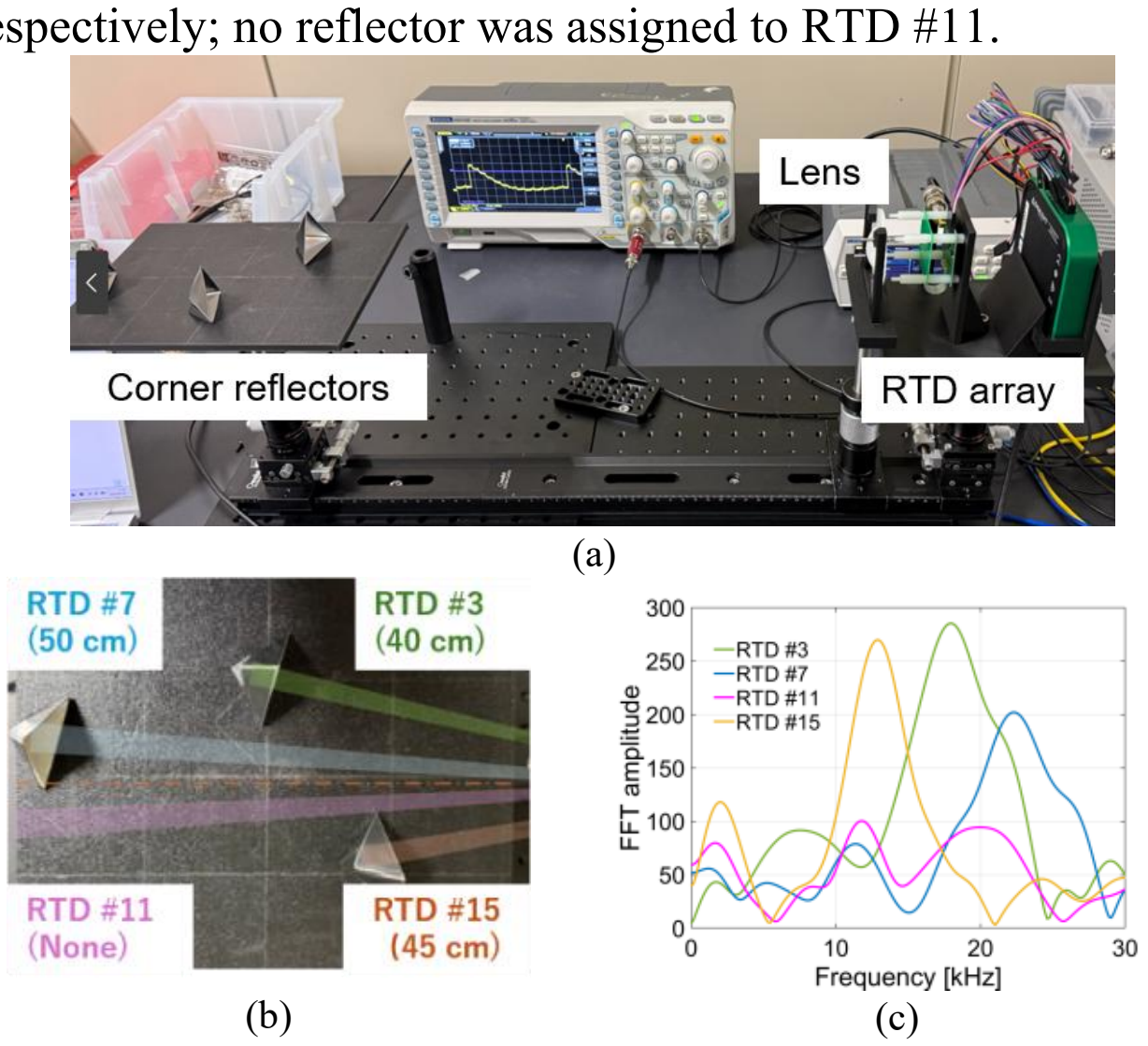

Fig. 7. (a) Experimental setup for sequential multidirectional radar operation. (b) Corner reflectors assigned to RTDs #3, #7, and #15 at 40, 50, and 45 cm, respectively; no reflector was assigned to RTD #11. (c) Range-detection results for the three corner reflectors.

As shown in Fig. 7(c), the radar beat frequency was estimated by applying an FFT to the RTD self-mixing signal, confirming multidirectional range estimation using sequentially addressed RTD elements. Because the RTDs operate at different bias voltages and do not have identical frequency excursions or chirp slopes, each selected element requires its own frequency-to-range calibration.

To address this issue, each radar sensor was calibrated using measurements at known target ranges, as shown in Fig. 8. For each RTD, the dominant beat frequency was extracted from the FFT spectrum after the same preprocessing procedure was applied to all measured signals. An RTD-specific linear calibration model was obtained by least-squares fitting and used to convert the measured beat frequency into range, compensating for device-dependent variations in the effective chirp slope and fixed processing offsets. The ranges measured using RTDs #3, #7, and #15 were 40.9, 54.7, and 44.2 cm, respectively, compared with reference ranges of 40, 50, and 45 cm. These measurements demonstrate sequential ranging in selected directions using the common bias and readout circuit.

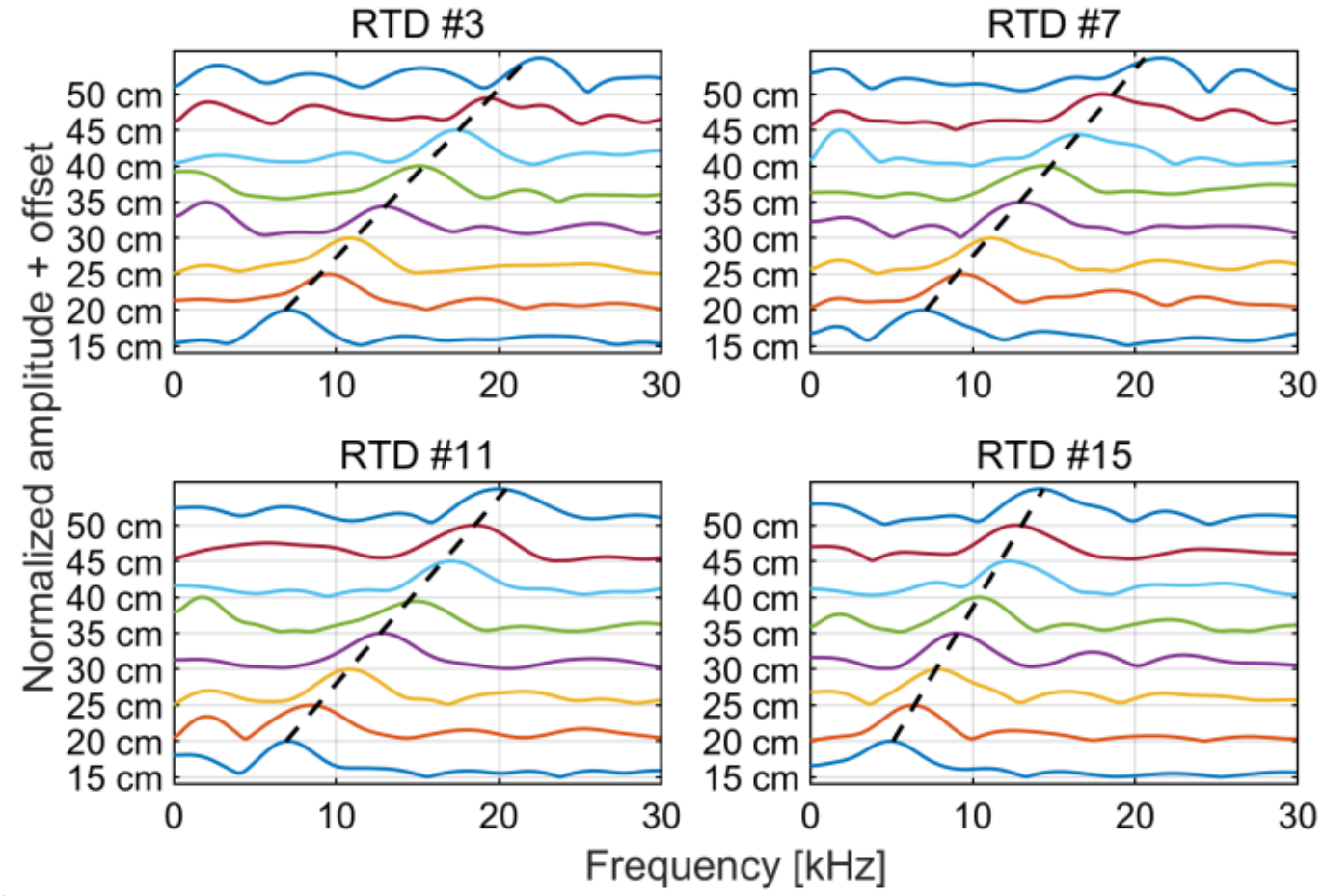

Fig. 8. Range-calibration data for the four RTD radar sensors. The dashed line indicates the linear fit to the calibration data.

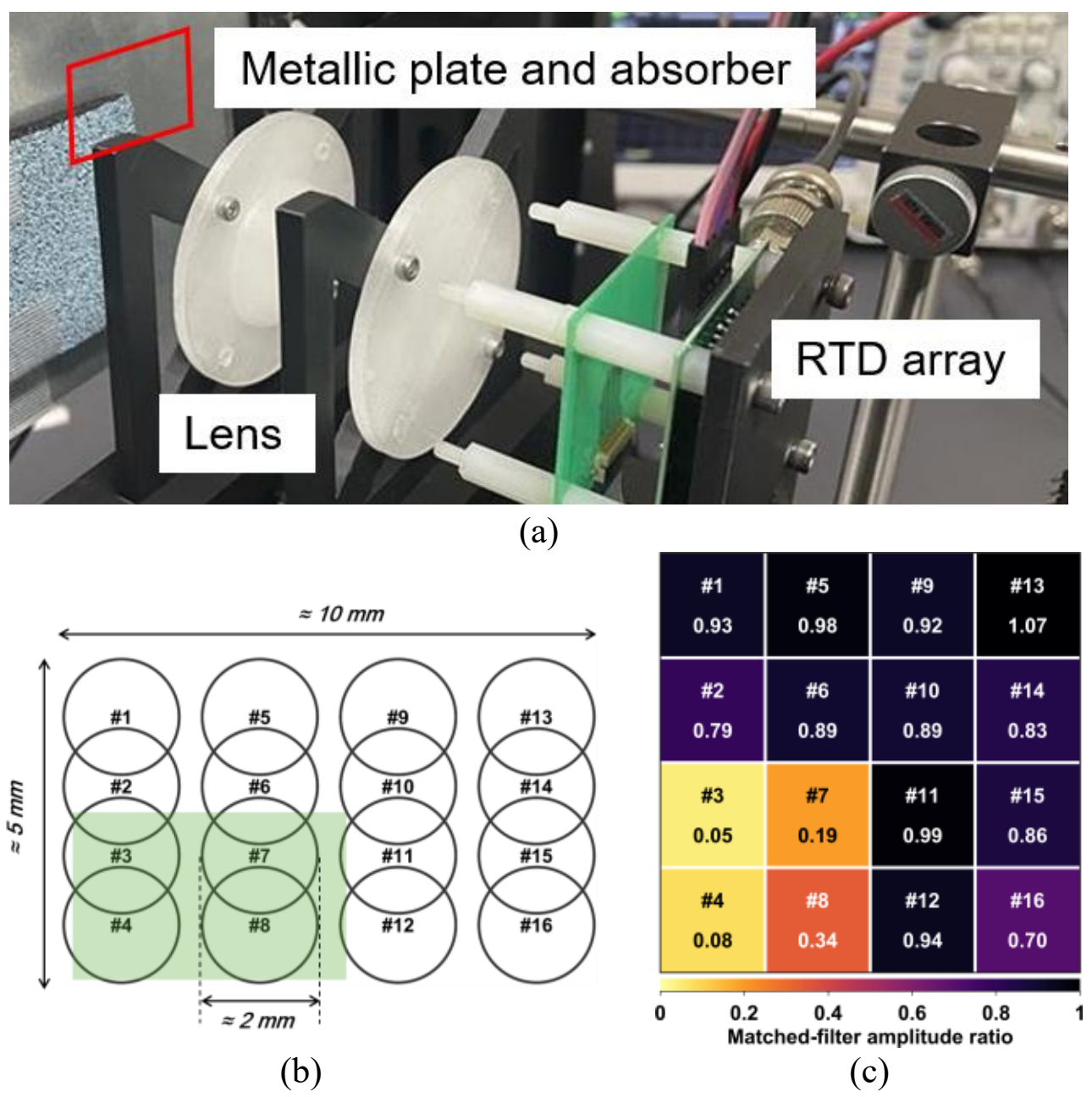

Fig. 9. (a) Experimental setup of the 16-pixel THz imager. (b) Projected beam-spot distribution in the imaging plane; the green shaded regions indicate the absorber positions. (c) Preliminary 16-pixel relative reflection-contrast image obtained using the 4 × 4 RTD monostatic array.

### B. Preliminary 16-Pixel THz Imaging Demonstration

Because each RTD element can operate as a THz transceiver, the proposed 4 × 4 array was also evaluated as a 16-pixel THz imager using the two-lens projection system shown in Fig. 9(a). The spatial distribution of the array was projected onto an approximately 10 mm × 5 mm target area, with a representative beam-spot diameter of approximately 2 mm, as illustrated in Fig. 9(b). A piece of low-reflectivity absorber was placed in front of a metallic plate to suppress the reflected signal locally. A template-matching method was applied to each pixel, and the aligned absorber-to-metal-plate amplitude ratio was used as the pixel response, as shown in Fig. 9(c). The reduced response in the absorber-covered region demonstrates spatial reflection contrast across the 16 sequentially acquired pixels. The ratio is a relative signal-amplitude measure rather than an absolute reflectivity, and the representative spot size is not used here as a quantitative resolution claim. This experiment is intended as a system-level proof of concept rather than a quantitative evaluation of imaging performance.

## Acknowledgment

This work was partially supported by the BOOST, JST (JPMJBY25A1).

## IV. Conclusion

This letter presented a 4 × 4, 350-GHz-band RTD monostatic radar array enabled by RTD self-mixing. Each RTD serves as both a tunable oscillator and a self-mixing detector, while low-frequency switches connect the selected pixel to a common baseband control and readout circuit. A shared 3D-printed telecentric lens increases the effective aperture and maps the array elements to spatially separated sensing directions. Experiments verified the operation and individual addressing of all 16 elements and demonstrated sequential multidirectional ranging using four selected pixels together with preliminary 16-pixel THz imaging. These results demonstrate an electronically addressable array architecture that avoids a separate THz receiver chain for each pixel.